\documentclass[prl,twocolumn,floatfix,notitlepage,superscriptaddress,longtable,nofootinbib]{revtex4-2}
\usepackage{amsmath}
\usepackage{lipsum} 
\usepackage{verbatim}
\usepackage{epsfig}
\usepackage{subfigure}
\usepackage{graphicx}
\usepackage{enumitem}
\usepackage{amsfonts}
\usepackage[final]{showlabels}
\usepackage{enumitem}
\usepackage[figuresright]{rotating}
\usepackage{amssymb}
\usepackage{amsmath}
\usepackage{psfrag}
\usepackage{mathtools}
\usepackage[export]{adjustbox}
\usepackage{xr}
\usepackage{bm}
\usepackage[colorlinks,linkcolor=blue,anchorcolor=blue,citecolor=blue,urlcolor=blue]{hyperref}
\usepackage[version=4]{mhchem}
\usepackage[svgnames]{xcolor}

\def\be{\begin{equation}} \def\ee{\end{equation}}
\def\bea{\begin{eqnarray}} \def\eea{\end{eqnarray}}

\def\bpm{\begin{pmatrix}} \def\epm{\end{pmatrix}}

\definecolor{Qicolor}{RGB}{3, 136, 252}

\makeatletter
\newcommand*{\balancecolsandclearpage}{%
	\close@column@grid
	\clearpage
}
\makeatother

\begin{document}

	\title{Decoherent histories with(out) objectivity in a (broken) apparatus}
	
	\author{Beno\^it Fert\'e}
	\affiliation{Universit\'e Paris-Saclay, CNRS, LPTMS, 91405, Orsay, France}
	\affiliation{Laboratoire de Physique de l'\'Ecole normale sup\'erieure, ENS, Universit\'e PSL, CNRS, Sorbonne Universit\'e, Universit\'e Paris Cit\'e, F-75005 Paris, France}
	
	\author{Davide Farci}
	\affiliation{Department of Physics and Astronomy, University of Florence, via Sansone 1, I-50019 Sesto Fiorentino (FI), Italy}
	
	\author{Xiangyu Cao}
	\affiliation{Laboratoire de Physique de l'\'Ecole normale sup\'erieure, ENS, Universit\'e PSL, CNRS, Sorbonne Universit\'e, Universit\'e Paris Cit\'e, F-75005 Paris, France}

	\date{\today}
	
	\begin{abstract}
		We characterize monitored quantum dynamics in a solvable model exhibiting a phase transition between a measurement apparatus and a scrambler. We show that approximate decoherent histories emerge in both phases with respect to a coarse-grained extensive observable. However, the apparatus phase, where quantum Darwinism emerges, is distinguished by the non-ergodicity of the histories and their correlation with the measured qubit, which selects an ensemble of preferred pointer states. Our results demonstrate a clear distinction between two notions of classicality, decoherent histories and environment-induced decoherence. 
	\end{abstract}
	\maketitle
	
	How classical reality emerges from quantum mechanics is a fascinating yet ambivalent question: the term ``classical'' is semantically overloaded. The goal of this Letter is to distinguish the notion of classicality in two influential approaches: environment-induced decoherence, and decoherent histories, in a situation where their distinction is demonstrable and instructive.
	
	The decoherence program~\cite{zeh-71,zeh-73,zurek-deco-review,zeh-review} is fundamentally motivated by a measurement problem, namely how to determine ``what is being measured'' when describing measurement as a unitary interaction between the system and the apparatus~\cite{vonneumann-book}. The key idea is that the apparatus (more generally, the environment) has many degrees of freedom, which allow to select a set of preferred ``pointer states'', through the proliferation of classical records~\cite{zurek-81,zurek-83}. In decoherence, and its refinement quantum Darwinism~\cite{ollivier-poulin,blume-kohout-zurek,zurek-QD,zurek-review,unden19-darwin-exp,girolami,mondani-darwism}, classicality is defined as \textit{objectivity}~\cite{Korbicz-prl,le-olaya-pra,le-prl19,Korbicz-rev}, the redundant accessibility of records.
	
	
	Meanwhile, the decoherent histories approach~\cite{griffiths-coherent-histories,griffiths-93-prl,gellmann-hartle-prd,dowker-halliwell-prd,omnes-review,halliwell-review} aims to describe classicality \textit{intrinsically}, solely from the multi-time correlation of an isolated quantum system under unitary evolution. Classicality emerges if the quantum evolution can be well approximated by a stochastic sum over an ensemble of trajectories (the decoherent histories) in some configuration space, with vanishing interference between them. 
	
	\begin{figure}
		\centering\includegraphics[scale=.9]{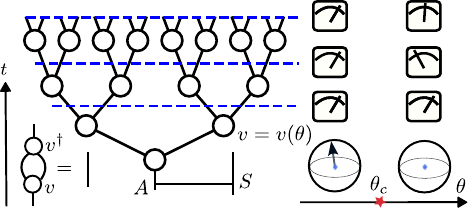} 
		\caption{A dynamically expanding tree model where every line is a qubit and every node is an isometry $v = v(\theta)$~\eqref{eq:vdef}. The root qubit $A$ is entangled with the qubit $S$~\eqref{eq:Psi}. The model has a transition at $\theta_c$~\eqref{eq:threshold}; it is an apparatus measuring $S$  if $\theta < \theta_c$, and a scrambler if $\theta > \theta_c$. In both phases, coarse-grained monitoring of the model~\eqref{eq:coarse},\eqref{eq:Km} yields decoherent histories~\eqref{eq:decay} (Fig.~\ref{fig:Delta}). In the encoding phase, the histories are Ornstein-Uhlenbeck like~\eqref{eq:Emtmt} and uncorrelated with $S$. In the apparatus phase, the histories are non-ergodic~\eqref{eq:freezinglaw} and selects an ensemble of pointer states of $S$ (see also Fig.~\ref{fig:pointer}). } \label{fig:tree}
	\end{figure}
	Despite the intrinsic nature of classicality à la decoherent histories, many works have investigated decoherent histories in open quantum systems, motivated by the relation with the decoherence program~\cite{finkelstein,saunders-93}. Some authors identify decoherent histories in the system (with the environment traced out) in the Markovian limit~\cite{zurek-dhc-prd,zurek-dhc-ptp}, and make connection with quantum trajectories~\cite{diosi-dh-qsd,yu-decoherence,brun-jump-deco,brun-trajectory-deco}. Others treat the system and the environment as a whole, and derive decoherent histories (or a related branching structure) from objectivity~\cite{riedel-consistent-QD,riedel-classical-branch,Touil2024branchingstatesas}. 
	This extensive literature fueled a folklore belief that decoherent histories require objectivity. This belief was challenged by recent works~\cite{strasberg-scipost,dhc-prx,wang-strasberg} showing evidence for (approximate) decoherent histories with respect to a coarse-grained and slowly varying observable, in an isolated macroscopic system under chaotic evolution. Note that chaotic dynamics is known to scramble information, making it inaccessible for all practical purposes~\cite{deutsch,srednicki,rigol-review}, which is the opposite of objectivity~\cite{riedel2012,duruisseau}. Hence it seems that decoherent histories can exist independently of objectivity~\cite{gemmer-16-markov,gemmer-steigeweg-14,classicality-thermalization}. However the mechanisms appear entirely different, and the relation between the two notions of classicality await clarification. 
	
\begin{figure}
		\centering\includegraphics[scale=.9]{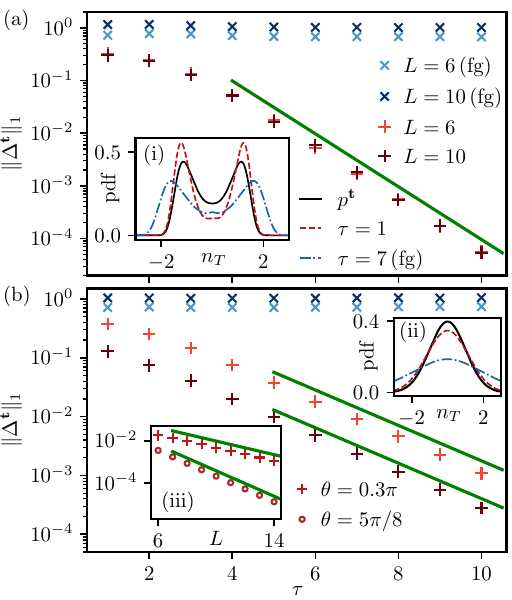}
		\caption{Decoherent histories probe $\Vert \Delta^{\mathbf{t}} \Vert_{1}$ with $\mathbf{t} = \{L+ \tau\}$, as function of $\tau$ (measurement starting time), for (a) $\theta = 0.15\pi < \theta_c$ and (b) $\theta = 0.3 \pi > \theta_c$.  Approximate decoherent histories emerge for coarse-grained measurements ($\Gamma = 0.1$) and $\tau \gg 1$, but not for $\tau \sim 1$ or for fine-grained measurements (fg, $\Gamma \eta_t = 0.1$). Insets (i, ii) show the pdf $p^{\mathbf{t}}$ (no third-party) and marginal distributions in non-decoherent scenarios ($T = 11$). Straight lines show decay rate of $ \Vert \Delta^{\mathbf{t}} \Vert_{1} $~\eqref{eq:decay}, and \eqref{eq:decayrate} in \cite{supp}. (a) For $\theta < \theta_c$, $ \Vert \Delta^{\mathbf{t}} \Vert_{1} \sim \Gamma^{-2} 2^{2\tau (x-1)}$ with no $L$ scaling. (b) For $\theta > \theta_c$, : $  \Vert \Delta^{\mathbf{t}} \Vert_{1} \sim  \Gamma^{-2}  2^{-T} \max(1, 2 \cos \theta)^{2L}$. Inset (iii): $L$ dependence is shown for two $\theta > \theta_c$ ($\tau = 6$). } \label{fig:Delta}
	\end{figure}

\begin{figure}
		\centering  \includegraphics[scale=.95]{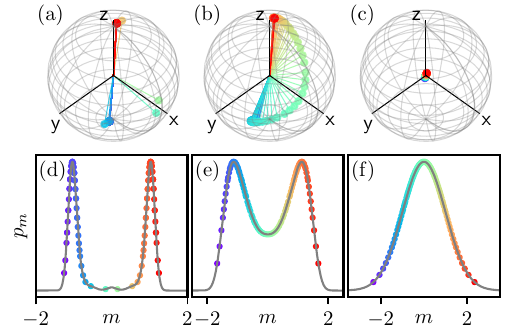} 
		\caption{Distribution of pointer states and marginal outcome distribution $p_m$ for $\theta = 0.05\pi < \theta_c$ (a,d), $\theta = 0.15\pi < \theta_c$ (b,e) and $\pi = 0.3 \pi > \theta_c$ (c,f). The color code relates the outcome to the condition state of $S$ inside the Bloch sphere. The points are distributed uniformly with respect to the probability. ($t = 20$ and $\Gamma = 0.05$.) }	\label{fig:pointer} 
	\end{figure}

	In this Letter, we characterize decoherent histories in a solvable model~\cite{fertecaoprl,fertecaopra} where objectivity can be turned on or off: the model has a transition between a functioning apparatus and a scrambler (Fig.~\ref{fig:tree}). We will show that decoherent histories emerge in both phases with respect to a coarse-grained observable, the coarse-graining being the essential common mechanism (Fig.~\ref{fig:Delta}). Nevertheless, we shall distinguish the phases by the (non)-ergodicity of the histories, and their correlation with the measured qubit, described in terms of a ``pointer state ensemble'' that we shall introduce (Fig.~\ref{fig:pointer}).

	\noindent\textbf{Model and objectivity threshold.} Consider one of the simplest unitary models of measurement~\cite{vonneumann-book}, where both the apparatus $A$ and the system $S$ are a qubit. Initially, they are disentangled and the apparatus is always in a ``ready'' state $| 0 \rangle_A$. The measurement unitary is such that $| i \rangle_S | 0 \rangle_A  \mapsto | i \rangle_S | i \rangle_A$, for $i = 0, 1$, so that measuring the superposition state  $( | 0 \rangle_S + |1 \rangle_S) / \sqrt{2} $ results in a maximally entangled pair:
	\begin{equation} \label{eq:Psi}
		| \Psi \rangle =  ( | 0 \rangle_S | 0 \rangle_A +  | 1 \rangle_S | 1 \rangle_A ) / \sqrt{2} \,.
	\end{equation}
	This is a situation where the aforementioned measurement basis problem is the most acute: it is impossible to know in which basis the measurement took place~\cite{zurek-81,zurek-83}. This issue is to be fixed by the presence of an environment or a more complex apparatus model, as we define now.
	
	In our model, $| \Psi \rangle$ is the initial state of a cascade process involving more and more qubits, bridging micro- and macroscopic realms~\cite{Ryan-onion,collision-model,campbell}. Starting with the qubit $A$, at each time step, every existing qubit interacts with a new qubit (in a pure state), so that there are $N_t = 2^t$ qubits after $t$ steps. During all this, $S$ remains intact. Such an inflationary dynamics is known as ``concatenation'' in the error correction code literature~\cite{concatenate,gullans-concatenate}, and often represented diagrammatically by a tree, see Fig.~\ref{fig:tree}~\cite{nahum21,feng2022measurement}. Every node corresponds to an isometry that outputs two qubits from one, $v: \mathbb{C}^2 \to \mathbb{C}^2 \otimes  \mathbb{C}^2$, such that $v^\dagger v = I$.  The evolution from time step $t$ to $t+1$ is given by $V_{t + 1, t}  = v^{\otimes 2^{t}}$. We also denote $V_{t,s} := V_{t, t-1} V_{t-1, t-2}  \dots V_{s+1, s}$ for $t \ge s$.  We will focus on the following concrete family of models:
	\begin{equation}\label{eq:vdef}
		v = \sum_{k = 0, 1} \left(e^{ - i X  \theta / 4 } | k \rangle \right)^{\otimes 2}  \langle k |  e^{ - i X  \theta  / 4 } \,,
	\end{equation}
	where $X$ is the Pauli-$x$ matrix, and $\theta \in (0, \pi / 2)$ is the tuning parameter.  Our model is equivalent to a quantum circuit with non-local control gates~\cite{supp}, similar to those realized by coherently transportable neutral atoms~\cite{bluvstein-22}.
	
	When $\theta = 0$, our model realizes a repetition code, which clearly has objectivity: every single apparatus qubit has perfect classical correlation with the system qubit. Increasing $\theta$ perturbs the code, and, as we shall see, destroys the objectivity beyond some threshold $\theta_c$.  To determine $\theta_c$, we consider the correlation between $S$ and an extensive observable of the apparatus, the total magnetization
	\begin{equation} \label{eq:coarse}
		\mathcal{Q}_t = \sum_{j=1}^{N_t} Z_j . 
	\end{equation}
	where $N_t = 2^t$ and $Z_j$ acts on the $j$-th qubit at time $t$. At $\theta = 0$, $\mathcal{Q}_t $ is also perfectly correlated with $Z_S$, the $Z$ operator on $S$. Exploiting the tree structure (see below and \cite{supp} for methods), we calculate the correlation between $O'_S$ (for any $O'$) and $	\mathcal{Q}_t$ for any $\theta$. We find the following asymptotic behavior (that is, the dependence on $t$ in the $t \to \infty$ limit):
	\begin{equation} \label{eq:onepoint}
		\mu_t := \langle  Z_S    V^\dagger_{t,0}  \mathcal{Q}_t  V_{t,0}   \rangle \sim  N_t^{1 - x_1}  
	\end{equation}
	where $ \left< [\dots ]\right> = \left< \Psi |  [\dots] | \Psi \right>$, and the exponent is
	\begin{equation} \label{eq:Delta}
		x_1 = -\log_2 (\cos \theta) .
	\end{equation}
	We also compute the scaling of the second moment of $\mathcal{Q}_t$:
	\begin{align} 
		\eta^2_{t} := & \langle V^\dagger_{t,0} \mathcal{Q}_t^2 V_{t,0} \rangle = \langle V^\dagger_{t,0}  \sum_{ij}  Z_i Z_j V_{t,0} \rangle  \nonumber \\
		\sim &  \begin{cases}
			N_t & x_1 > 1/2 \,,\,  \\
			N_t^{2(1-x_1)} & x_1 < 1/2 \,. \label{eq:twopoint}
		\end{cases}
	\end{align}  
	The threshold corresponds to $x_1 = 1/2$ or 
	\begin{equation}\label{eq:threshold}
		\theta_c = \pi/4 \,.
	\end{equation} 
	When $\theta < \theta_c$, $x_1 < 1/2$, $\mu_t \sim \eta_t$, the correlation between $\mathcal{Q}_t $ and any operator on $S$ is comparable to its fluctuation amplitude. Moreover, \eqref{eq:onepoint} and \eqref{eq:twopoint} hold as well if  $\mathcal{Q}_t$~\eqref{eq:coarse} is replaced by the total magnetization of a fixed fraction $f N_t$ of apparatus qubits~\cite{supp}. Thus, objectivity prevails: information about $S$ can be obtained from small fractions of the apparatus. We call $\theta < \theta_c$ the ``apparatus phase''.  When $\theta > \theta_c, x_1 >  1/2$, the correlation is vanishing compared to the noise, $\mu_t \ll \eta_t$, and observing $\mathcal{Q}_t $ yields no information about the qubit. This is the ``encoding phase''. Based on previous work on a very similar model~\cite{fertecaopra}, we expect that the Holevo quantity between an apparatus fraction and $S$ vanishes if $\theta > \theta_c$ as $t\to\infty$. In other words, information about $S$ is inaccessible to other observables as well, not just $\mathcal{Q}_t$. 
	
	We remark that our expanding tree model is a special case of the multi-scale entanglement renormalization Ansatz  tensor networks \cite{vidal-mera-0,vidal-mera,evenbly-vidal,banul-review-TN},  which are known to generate scale invariant states. In that context, the exponent $x_1$ above is the lightest scaling dimension. Interestingly, the threshold phenomena at $x_1 = 1 / 2$ appeared in various recently studied scale invariant ``inference'' problems~\cite{garratt,patil2024highly,nahum2025bayesian}.

	\noindent\textbf{Decoherent histories.}  We now turn to decoherent histories in our model. Our approach builds upon that of \cite{dhc-prx}, which we first review. The standard decoherent histories formalism applied to our model would start with a family of projectors $\{ K^t_m \}$ that sum to $\mathbf{1}$ for each time step $t \in [\tau, T]$. Then one defines, for $\vec{m} = (m_\tau, \dots, m_T)$, the state
	\begin{equation}\label{eq:mstate-def}
		| \vec{m} \rangle :=    K^T_{m_T} V_{T, T-1}  \dots V_{\tau + 1, \tau} K^\tau_{m_\tau} V_{\tau, 0} | \Psi \rangle  \,.
	\end{equation}
	Now, the norm of these states 
	\begin{equation} \label{eq:pm-def}
		p_{\vec{m}} = \langle  \vec{m}  | \vec{m} \rangle
	\end{equation}
	are also the outcome probability distribution if one measures the apparatus at each time step projectively using $\{K_m^t\}$. The (exact) decoherent histories condition (DHC) is for these states to be orthogonal:
	\begin{equation}
		\forall \, \vec{m} \ne \vec{m}' \,,\,	\left< \vec{m} | \vec{m}'  \right> = 0   \quad \text{(DHC)}
	\end{equation}
	A consequence of the DHC is that, if measurements are only performed at any time subset $ t \in \mathbf{t}  = \{t_1 < t_2 \dots  < t_k\} \subset \{\tau, \dots, T\}$, the outcome distribution
	\begin{equation}
		p^{\mathbf{t}}_{\vec{n}}  = \Vert   K_{n_k}^{t_k}  V_{t_k, t_1}  \dots   V_{t_2, t_1} K_{n_1}^{t_1}  V_{t_1, 0}   | \Psi \rangle  \Vert^2
	\end{equation}
	will be equal to the marginal of $p_{\vec{m}}$ to the time set $\mathbf{t}$:
	\begin{equation}\label{eq:diff}
		\text{DHC} \implies	\Delta^{\mathbf{t}}_{\vec{n}} := p^{\mathbf{t}}_{\vec{n}} - \sum_{\vec{m}\vert_\mathbf{t} = \vec{n}}	 p_{\vec{m}}  = 0 \,,
	\end{equation}
	where $ \vec{m}\vert_\mathbf{t} = (m_{t_1}, \dots, m_{t_k})$. Without DHC, $\Delta^{\mathbf{t}}_{\vec{n}} $ has no reason to vanish. In fact, outcome probability differences like $ \Delta^{\mathbf{t}}_{\vec{n}} $ are essentially the only operational way to detect coherence. So it is reasonable to use the vanishing of \eqref{eq:diff} (for all $\textbf{t}, \vec{n}$) as a probe of emergent approximate decoherent histories, as did the authors of Ref.~\cite{dhc-prx}.  
	
	In the sense of \eqref{eq:diff},  classicality à la decoherent histories amounts to non-disturbance by third-party measurements (which is also part of macro-realism~\cite{leggett-garg}). Namely, classicality emerges if the \textit{existence} of measurements in $t \notin \mathbf{t}$ does not affect the outcome distribution of the ones in $t \in \mathbf{t}$. This is a reasonable way to define ``classical'' since only classical systems admit ideal measurements that reveal their state without changing it. 
	
	Given the above observation, we shall replace projective measurements by weak measurements described by smeared Kraus operators~\cite{Wiseman_Milburn_2009}, which are more suitable for a coarse-grained observable. Concretely,  we let
	\begin{equation} \label{eq:Km}
		K^t_m = \frac{1}{(2 \pi \Gamma^2)^{1/4}}  e^{- (\mathcal{Q}_t / \eta_t- m)^2 / (4 \Gamma^2)} \,, m \in \mathbb{R}, 
	\end{equation} 
	where $\Gamma > 0$ is $t$-independent. Namely, $K^t_m $ describes measuring the extensive observable~\eqref{eq:coarse} rescaled by its uncertainty~\eqref{eq:twopoint}, with $O(1)$ relative precision tuned by $\Gamma$.  The idea of measuring the apparatus may evoke the Wigner friend's paradox~\cite{Wigner1995}. However, here, the ``super-observer'' is just a mathematical probe of the intrinsic dynamics of the apparatus (see also ~\cite{gemmer-16-markov,gemmer-steigeweg-14,classicality-thermalization,cao2025planckian}).
	
	It turns out that approximate decoherent histories emerge provided $\tau \gg1$, that is, when the apparatus is macroscopic. We show in \cite{supp} that for any $\mathbf{t}$,
	\begin{equation} \label{eq:decay}
		\Delta_{\vec{n}}^{\mathbf{t}} = O((\Gamma \eta_\tau)^{-2}) \,, \tau \to \infty \,.
	\end{equation}
	Here we focus on the case $\mathbf{t} = \{T\}$, which has the most third-party measurements $t < T$ causing possible disturbance. Since the probe \eqref{eq:diff} involves a single variable, $n_T$, it is amenable to exact numerics. We evaluate the $L^1$ norm:
	\begin{equation}
		\Vert \Delta^{\mathbf{t}} \Vert_1 :=   \int \left| \Delta^{\mathbf{t}}_{n_T} \right|  d n_T  \,,
	\end{equation}
	whose vanishing indicates the emergence of decoherent histories. Some results are shown in Fig.~\ref{fig:Delta}. 
	We observe that $\Vert \Delta^{\mathbf{t}} \Vert_1$ decays exponentially with respect to $\tau$  as predicted \eqref{eq:decay} for fixed $L = T - \tau$ in both phases. With respect to $L \gg 1$, there is essentially no dependence in the apparatus phase, while in the encoding phase, $\Vert \Delta^{\mathbf{t}} \Vert_1$ also decays exponentially. So, for small $\tau$, $ \Vert \Delta^{\mathbf{t}} \Vert_1 $ vanishes at large $T$ in the encoding phase and not in the apparatus phase. These results are explained as follows. Third-party measurements at the early, microscopic, stage will disturb the apparatus dynamics in both phases. In the apparatus phase, the disturbance remains accessible at late time. In the encoding phase, the disturbance is only visible at early time and becomes scrambled later.
	
	The emergence of decoherent histories crucially relies on coarse-graining, that is, measuring an extensive quantity with $\Gamma = O(1)$ \textit{relative} precision, see~\eqref{eq:Km}. Had we let $\Gamma \sim O(1/\eta_t)$, $\Vert \Delta^{\mathbf{t}} \Vert_1  \not\to 0$ in any limit (see blue crosses in Fig.~\ref{fig:Delta}).  Such a measurement with $O(1)$ \textit{absolute} precision would disturb the dynamics even in a macroscopic system, revealing its underlying quantum nature.  Ref.~\cite{brukner-leggett-garg} also pointed out the importance of measurement imprecision for classicality in macroscopic systems, in terms of the Leggett-Garg inequality~\cite{leggett-garg}. Interestingly, in our model, this inequality can be violated at arbitrarily late times in both phases, by operators of type $Q_t = (a X_j + b Y_j + c Z_j)^{\otimes 2^t}, Q_t^2 = \mathbf{1}$~\cite{supp}.  Again, such an observable reveals the many-body coherence of the model inaccessible to coarse-grained ones. 
	
	\noindent\textbf{Histories statistics.} Although decoherent histories emerge at $t \ge \tau \gg 1$ independently of objectivity, they have distinct statistics $p_{\vec{m}}$~\eqref{eq:pm-def} in the two phases. In the encoding phase ($x_1 > 1/2$), the time sequence $(m_t)$ is a Gaussian process with finite temporal correlation:
	\begin{equation} \label{eq:Emtmt}
		\mathbb{E}[m_t] = 0,	\mathbb{E}[m_t m_{t'}] \sim e^{- \kappa  | t - t' |  } \,,\,  \kappa = (x_1- 1/2 ) \ln 2 
	\end{equation}
	for $|t-t'| \gg 1$. The correlation time $\kappa^{-1}$ is finite (it only diverges near the threshold $x_1 = 1/2$). To understand this intuitively, recall that the extensive observable $\mathcal{Q}_{t \gg 1}$ reveals no initial information (stored in $S$); since the expanding dynamics is self-similar, information about the system at any time $t$ becomes inaccessible at $t' - t \gg 1$ as well.
	
	Meanwhile, the apparatus phase ($x_1 < 1/2$) is characterized by histories with the following probabilistic law: 
	\begin{equation} \label{eq:freezinglaw}
		m_t \stackrel{\text{in law}} = \mathcal{M} + \Gamma \, \xi_t
	\end{equation}
	where $\xi_t$ are independent identically distributed standard Gaussian and $\mathcal{M}$ is a random variable with the same distribution as $\mathcal{Q}_t / \eta_t$ in the long time limit (this distribution is non-Gaussian, see Fig.~\ref{fig:Delta}):
	\begin{equation}\label{eq:lawofM}
		\forall f  \,,\,	\overline{f(\mathcal{M})} =  \lim_{t \to\infty}  \langle  V_{t,0}^\dagger f(\mathcal{Q}_t / \eta_t) V_{t,0} \rangle. 
	\end{equation}
	Namely, in every experimental run, the outcome $m_t$ becomes independent of $t$ up to an $O(1/\Gamma)$ noise, yet this constant value is random and varies at each run. Such behavior is akin to that of an apparatus pointer that freezes around a random value. The histories are non-ergodic: the statistical average over many runs differs from the time average over a single run. The correlation time is infinite: for any $t, t'$, $m_t$ and $m_{t'}$ are correlated. This is because $m_{t, t'}$ is correlated with $\mathcal{Q}_{t,t'}$ (respectively),  and, as we have seen above, they are both correlated with the qubit $S$.
	
	
	\noindent\textbf{Pointer states ensemble.} We come back to the question of what information on the measured qubit $S$ is inferred from an apparatus history. This can be described by the ensemble of the conditional state $\rho_{\vec{m}}$ of $S$ after a coarse-grained monitoring [using \eqref{eq:Km}] of the apparatus, weighted by Born's rule $p_m$~\cite{fertecaopra,gullans-huse-prl,deep-therm,deep-therm-ho}, which we call the pointer states ensemble.  This ensemble is different from and ``softer'' than the usual definitions of pointer states~\cite{zurek-81,zurek-dhc-ptp,pointer-korbiz}. In general, it is made of mixed states that do not form a basis. Yet, they constitute a ``decomposition of unity'' like a basis: $\sum_{\vec{m}} p_{\vec{m}} \rho_{\vec{m}} = I$~\cite{fertecaopra}. Moreover, it captures an imperfect form of apparatus super-selection, as we now illustrate.
	
	One may argue~\cite{supp} that a full history of $t \in [\tau, T], \tau \gg1$ infers no more information than a single-time measurement using \eqref{eq:Km} at $t \gg 1$, so we focus on the latter. Eq.~\eqref{eq:Psi} implies that the conditional density matrix of $S$ is 
	\begin{equation}
		\rho_{m}= (V_{t,0}^\dagger (K^t_m)^\dagger K^t_m V_{t,0})^{\mathsf{T}} / (2p_m)
	\end{equation} 
	where $\mathsf{T}$ is the transpose. This can be efficiently computed, and has a well-defined $\Gamma \to 0, t \to \infty$ limit. Fig.~\ref{fig:pointer} shows the ensemble of $\rho_m$ inside the Bloch sphere. The ensemble concentrates more and more on $(I \pm Z) / 2$ as $\theta \to 0$. This is the limit of a perfect apparatus that super-selects the pointer state basis $\{ |0 \rangle, |1\rangle \}$. For $0 < \theta  <\theta_c$, we have an imperfect apparatus. There is a nonzero probability that the apparatus' freezes at $m \sim 0$, while the qubit does not ``collapse'' towards either $| 0 \rangle$ or $| 1\rangle$, but remains a superposition, $\rho_m \approx I + X$. In general, $\rho_m$ is mixed, meaning that the ``collapse'' is incomplete. Nevertheless, there is a temporal consensus~\cite{touil2025consensus} by virtue of \eqref{eq:freezinglaw}: super-observers at different $t$ will agree with each other on what the apparatus did. In the encoding phase, $\rho_m \to  I / 2$: there is no meaningful pointer state.
	
	\noindent\textbf{Conclusion.} We distinguished two notions of classicality. Defined via decoherent histories, its emergence in a macroscopic object relies only on coarse-graining and is likely generic. Yet, classicality à la quantum Darwinism is about the particular macro-microscopic correlation as realized during a quantum measurement. An apparatus is a macroscopic object, but not every macroscopic object is an apparatus. This statement may sound obvious, yet is nontrivial to demonstrate: we only did it in a somehow unusual ``inflationary'' model. Indeed, it would be desirable to confront objectivity and decoherent histories in conventional lattice models beyond small sizes. There, stabilizing objectivity is nontrivial, and plausibly requires some form of symmetry breaking~\cite{riedel2012,giorgi,duruisseau,campbell}. Meanwhile, the implication of our study for the issue of classicality in cosmology~\cite{classicalty-cosmo} awaits exploration.
	
	\begin{acknowledgements}
		We thank Serge Florens and Vincent Vennin for useful discussion and collaboration on related topics.
		
		The data that support the findings of this article are openly available~\cite{data}.
	\end{acknowledgements}

	\bibliography{ref}
			
			\subsection{Quantum circuit realization}
			The ``inflationary'' quantum dynamics of our model can be represented as a quantum circuit with non-local control gates. Below is a diagram for $t = 3$:
			\begin{equation*}
				\includegraphics[width=\columnwidth]{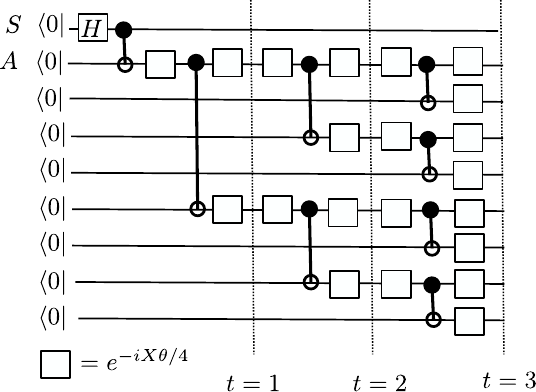}
			\end{equation*}
			Quantum circuits with non-local control gates have been realized, for instance by coherently transported neutral atoms. By making computational basis measurement at the end of the circuit, and post-processing the results, it is possible to probe, modulo finite-time effects, the objectivity-encoding threshold phenomenon as characterized by the correlation between $S$ and the rest of the qubits. 
			
			\subsection{Objectivity threshold and real-space RG}
			We detail the derivation of \eqref{eq:onepoint} and \eqref{eq:twopoint} in the main text and provide the exact prefactors. As we mentioned in the main text, the calculation is a simple application of the real-space RG idea in developed in the MERA tensor network context. 
			
			
			
			\noindent{\bf Scaling dimensions and OPE}.	 
			We first recall the definition of a scaling operator in the MERA tensor network context. For our purposes, it suffices to specialize to tree-MERA tensor networks. Then, the scaling operators act on one qubit, and there are four of them. We denote them by $O^{(0)}, \dots, O^{(4)}$. By definition, they satisfy the following:
			\begin{equation} \label{eq:Deltadef}
				v^\dagger ( O^{(a)} \otimes \mathbf{1}) v =  v^\dagger ( \mathbf{1} \otimes O^{(a)} )	v =  2^{-x_a}  O^{(a)}.
			\end{equation}
			where $x_a$ are the corresponding scaling dimensions. Using this definition, we can check explicitly that all the scaling operators and their scaling dimensions are listed in the following table:
			\begin{equation}
				\begin{tabular}{|c|c|c|} 
					\hline
					$a$ & $O^{(a)}$ & $x_a$ \\  \hline
					0		& $I$  & $0$ \\ 
					1	&  $Z \cos(\theta/2) + Y  \sin (\theta/ 2)$  & $-\log_2 \cos \theta$ \\ 
					2	&		$X$ & $ \infty$ \\ 
					3	&	$Z \sin(\theta/2) + Y \cos(\theta/2)$  & $ \infty$  \\  \hline
				\end{tabular}
			\end{equation}
			In particular, the $0$-th scaling operator is the identity. $O^{(1)}$ is the lightest non-identity scaling operator. The last two operators with infinity scaling dimension are annihilated by upon ``coarse-graining'', $v^\dagger (O^{(2,3)} \otimes I) v = 0 $ and $v^\dagger (I \otimes O^{(2,3)}) v = 0 $. 	
			In terms of the scaling operators, the local $Z$ operator is decomposed as follows:
			\begin{equation} \label{eq:Z-decomp}
				Z = O^{(1)} \frac{\cos(\theta/2)}{\cos\theta}  - O^{(3)} \frac{\sin(\theta/2)}{\cos\theta} \,.
			\end{equation}
			
			We also recall the notion of operator product expansion (OPE) coefficients between non-identity operators. They are defined as 
			\begin{equation}
				v^\dagger ( O^{(a)} \otimes  O^{(b)}) v = \sum_c C_{ab}^c O^{(c)}, a, b > 0.
			\end{equation}
			The most important OPE is that between $O^{(1)}$ and $O^{(1)}$. The OPE coefficients are:
			\begin{equation}
				C_{11}^0 =  \cos (\theta)^2 , C_{11}^1 = 0, C_{11}^2 = - \sin (\theta)^2, C_{11}^3 = 0.
			\end{equation}
			The OPE coefficients involving the other scaling operators with infinite dimension will not be useful for what follows, so we shall not enumerate them. However it is worth noting that $C_{22}^2 = 1$, or $v^\dagger (X \otimes X) v = X$. This means that the product operator $\prod_j X_j$ is conserved by the dynamics. We will exploit this fact when showing a violation of the Leggett-Garg inequality, see below.

			
			\noindent	\textbf{One point function}. We now compute the correlation between $\mathcal{Q}_t = \sum_j Z_j$ and an operator $O'$ acting on the qubit $S$. From the definition of scaling dimension and that of the many-body evolution operator $V_{t+1,t} = v^{\otimes 2^t}$, it follows that 
			\begin{equation}\label{eq:onepoint1}
				V_{t,0}^\dagger O_{j}^{(1)} V_{t,0} = 2^{-x_1 t} O^{(1)}
			\end{equation}
			where $O_{j}^{(1)} $ is $O^{(1)}$ acting on the qubit $j$ at time $t$, and $O^{(1)}$ on the right hand side acts on the tree root qubit $A$, as illustrated below (with $t = 3$): 
			\begin{equation*}
				\includegraphics[scale=1]{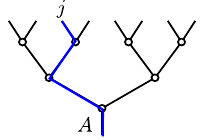}
			\end{equation*}
			Also, $V_{t,0}^\dagger O^{(a)}_{j} V_{t,0} = 0$ for $a = 2, 3$ and for any $t > 0$. From \eqref{eq:Z-decomp} and \eqref{eq:onepoint1} we have for any $t > 0$, and any operator $O'_S$ acting on the measured qubit $S$ (which is entangled with $A$),
			\begin{align} \label{eq:onepoint-app}
				\left< O'_S V_{0, t}^\dagger \mathcal{Q}_t V_{0,t} \right> =  \frac{\cos(\theta/2)}{\cos\theta}   \langle   O'_S  O^{(1)} \rangle 2^{t(1-x_1)} \,,
			\end{align}
			where the $\left< [\dots] \right> = \langle \Psi | [\dots] | \Psi \rangle$ is the average with respect to the initial entangled state on $A$ and $S$. This gives the scaling $\mu_t \sim 2^{t (1 - x_1)}$ stated in the main text.
			
			\noindent	\textbf{Two point function}. Next we calculate the two point correlation. For this we first claim the following:
			\begin{equation} \label{eq:twopoint0}
				V_{T, 0}^\dagger O_{i}^{(1)} O_{j}^{(1)}  V_{T, 0} = 2^{-2 x_1 r}  C_{11}^0  I + c X \,,\, i \ne j \,,
			\end{equation}
			where $r$ is the number of nodes between $i$ or $j$ and their common ancestor. Here is an illustrated example in a model with $t = 3$:
			\begin{equation*}
				\includegraphics[scale=1]{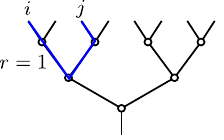} \,.
			\end{equation*}
			The operator dynamics evolves (backwards) $r$ times each operator independently (giving $2^{-2 x_1 r}$)  before fusing them into identity or $O^{(2)} = X$. The former possibility gives rise to the term $ 2^{-2 x_1 r}  C_{11}^0  I$ in \eqref{eq:twopoint0} since the identity then evolves trivially. The latter possibility gives rise to the term $cX$ where the coefficient $c$ does not vanish only if $r = t - 1$. Even in that case, that term will not contribute to the variance of $\mathcal{Q}_t$ since $\left< X \right> = 0$, so we will ignore it below.
			
			There are $2^{t + r}$ pairs of $i\ne j$ with the same distance $r$, for $r= 0, \dots, t-1$. Therefore 
			\begin{align} 
				\sum_{i\ne j}  \left< V_{T, 0}^\dagger O_{i}^{(1)} O_{j}^{(1)} V_{T, 0}  \right> &= 2^t \sum_{r=0}^{t-1} 2^{(1-2 x) r}  C_{11}^0   \nonumber \\
				&= 2^{t}  \frac{2^{t(1-2x)} - 1}{2^{1-2x} - 1} C_{11}^0   \label{eq:two-point-app}
			\end{align}
			Then, for $t > 2$ (to ignore the subleading scaling operators), we have 
			\begin{align*}
				&\sum_{i,j}  \left< V_{T, 0}^\dagger Z_i Z_j V_{T, 0} \right> = 2^t  + 2^{t}  \frac{2^{t(1-2x_1)} - 1}{2^{1-2x_1} - 1} C_{11}^0\frac{\cos(\theta/2)^2}{(\cos\theta)^2}   \,.
			\end{align*}
			The asymptotic behavior is the following, with a correction to scaling at the threshold $x_1 = 1/2$:
			\begin{equation} \label{eq:etat_app}
				\eta_t^2 \sim  \begin{dcases}
					\frac{\cos(\theta/2)^2 }{\cos (2\theta)} \,	2^{2t(1-x_1)}  & x_1 < 1/2 \\ 
					\frac{\cos(2\theta)  -\cos(\theta/2)^2  }{\cos (2\theta)}  \, 2^t & x_1 > 1/2  \\
					\cos(\theta/2)^2 \,	t 2^t	& x_1 = 1/2
				\end{dcases} 
			\end{equation}
			In what follows we will assume $x_1 \ne 1/2$.
			
			\noindent \textbf{Fractions and objectivity.}	We now discuss what happens if the coarse-grained operator $\sum_{j=1}^{2^t} Z_j$ is replaced by a sum over a fraction of the apparatus qubits,  $F \subset \{ 1, \dots, 2^t\}$,
			\begin{equation}
				\mathcal{Q}_F = \sum_{j \in F} Z_j \,.
			\end{equation}
			We shall denote by
			\begin{equation}
				f = |F| / 2^t
			\end{equation}
			the relative volume fraction of $F$. The correlation of $ $ with $O'_S$ can be computed in the same way as \eqref{eq:onepoint-app}:
			\begin{equation}
				\mu_{t,f} = \left< O'_S V_{0, t}^\dagger \mathcal{Q}_F  V_{0,t} \right> =  \frac{\cos(\theta/2)}{\cos\theta}   \left<   O'_S  O^{(1)} \right> 2^{t(1-x_1)} f \,.
			\end{equation}
			Compared to \eqref{eq:onepoint-app}, it has the same scaling law, and the pre-factor changes by $f$. Now, the second moment of $\mathcal{Q}_F $ depends on how $F$ is distributed with respect to the tree structure. Let us discuss two cases. First, let $F$ be a spatially compact  subtree, with $f = 2^{-t_0}$, as illustrated by the green boxes below (with $t_0 = 1$, $t = 3$):
			\begin{equation*}
				\includegraphics{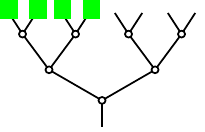}
			\end{equation*}
			Then eq.~\eqref{eq:two-point-app} will be modified as follows: $2^t$ is replaced by $2^{t-t_0}$, and the sum over $r$ will have an upper limit $r \le t - 1 - t_0$. As a result, the scaling laws \eqref{eq:etat_app} do not change, and the prefactor is modified by $f$-dependent factor as follows:
			\begin{equation}
				\eta_{t,f}^2 / \eta_{t, f= 1}^2  =  \begin{dcases}
					f^{2 (1-x_1)}   & x_1 < 1/2 \\ 
					f  & x_1 > 1/2
				\end{dcases}	 
			\end{equation}
			As a consequence, in the apparatus phase ($x < 1/2$), the signal to noise ratio has an asymptotic $f$ dependence $\mu_{t, f}  /   \eta_{t,f} \sim f^{x}$. It remains finite for any fixed relative fraction $f$. But it vanishes as $f \to 0$ unless the apparatus is perfect, $x = 0$. Hence the information about $S$ is accessible (with a given precision) to a finite number of  spatially compact fractions. 
			
			Second, let $F$ be a dilute fraction, uniformly distributed across the subtrees, with $f = 2^{-t_0}$, as illustrated below  (with $t_0 = 1$, $t = 3$):
			\begin{equation*}
				\includegraphics{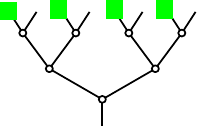}
			\end{equation*}
			Then eq.~\eqref{eq:two-point-app} will be modified as follows: the sum over $r$ will have an lower limit $r \ge t_0$, and there will be an extra $2^{-2t_0}$ factor. Again, the scaling laws \eqref{eq:etat_app} prevail, and the $f$-dependent prefactor is 
			\begin{equation}
				\eta_{t,f}^2 / \eta_{t, f= 1}^2  = f^2  \,,\, x_1 < 1/2 \,.
			\end{equation}
			(The case $x_1 > 1/2$ is more cumbersome and unimportant). Hence, in the apparatus phase, the signal-noise ration $\mu_{t, f} /\eta_{t,f} \sim 1$ has no $f$-dependence in the $t \to \infty$ limit. Hence some information about $S$ is accessible to an arbitrarily large number of dilute fractions in the late time limit. It is in this sense that the apparatus phase has emergent objectivity.
			
			\subsection{Decoherent histories: exact methods}
			We describe how to use the solvability of the tree model to study decoherent histories, both analytically and with efficient and exact numerics. 
			
			\noindent\textbf{Generality.} Our method works in the Heisenberg picture, that is, it focuses on operators rather than states. It also naturally treats the decoherent histories and their correlation with $S$ at the same time. We first define the time evolution and Kraus super-operators:
			\begin{equation}
				\mathcal{V}_{t',t}[\mathbf{Q}] =  V_{t,t'}^\dagger \mathbf{Q} V_{t,t'} \,,\,  \mathcal{K}^t_{m}[\mathbf{Q}] = (K^t_m)^\dagger \mathbf{Q} K^t_m  \,,
			\end{equation}
			Then consider the operator acting on $\mathbb{C}^2$ (the Hilbert space of $A$):
			\begin{equation}
				Q_{\vec{m}} :=	\mathcal{V}_{0,\tau}    \mathcal{K}^{\tau}_{m_{\tau}}   \mathcal{V}_{\tau, \tau+1}  \dots  \mathcal{K}^{T-1}_{m_{T-1}} \mathcal{V}_{T-1,T} \mathcal{K}^T_{m_T} [\mathbf{1}] \,.
			\end{equation}
			It is not hard to see using \eqref{eq:Psi} that the history probability is given by the rescaled trace:
			\begin{equation}
				p_{\vec{m}} =  \mathrm{tr}(Q_{\vec{m}}) / 2 \,,
			\end{equation}
			and the conditional density matrix of $S$ is 
			\begin{equation}
				Q_{\vec{m}} =  Q_{\vec{m}}^\mathsf{T}  / (2 p_{\vec{m}}) \,.
			\end{equation}
			In other words, up to a transpose, $Q_{\vec{m}}$ is the non-normalized conditional density matrix of $S$, and the normalization factor is the history probability.
			
			Now, applying a Hubbard-Stratonovich transform to each Kraus operator, 
			\begin{equation}
				K^t_m \propto \int d u  e^{- \Gamma^2 u^2 - i m u / \eta_t }  \prod_j e^{i u Z_j /  \eta_t},
			\end{equation}
			we may reduce the many-body operator dynamics into (nonlinear) transformations of a $2\times 2$ matrix $Q$. Indeed, in terms of the super-operators acting on $2\times2$ matrices,
			\begin{align} \label{eq:vk}
				&	\mathsf{v} [Q] := v^\dagger (Q \otimes Q) v  \,,\,  \\
				&	k^t_{u_t, v_t}[Q] := e^{i {u_t} Z / \eta_t } {Q} \, e^{-i  v_t Z  / \eta_t}
			\end{align}
			(do not confuse the number $v_t$ and the vector $\vec{v}$ with the isometry $v$), we have 
			\begin{align}\label{eq:pm-int}
				Q_{\vec{m}} \propto \int_{\vec{u}, \vec{v}} &   e^{ - \sum_t  (\Gamma^2 (u_t^2 + v_t^2)  + i (u_t- v_t) {m_t} )} \hat{Q}_{\vec{u}, \vec{v}} 
			\end{align}
			where $\vec{u} = (u_\tau, \dots, u_T), \vec{v} = (v_\tau, \dots, v_T) $, and $\hat{Q}_{\vec{u}, \vec{v}} $ is defined by a backward recursion:
			\begin{align}\label{eq:hatp-recurs}
				&\hat{Q}_{\vec{u}, \vec{w}} = (\mathsf{v})^{\tau-1} [Q_{\tau-1} ] \\ 
				& Q_{t-1}= \mathsf{v} k^t_{u_t, v_t} [Q_{t}] , 
				Q_T = I_2 \,. \nonumber
			\end{align}
			The proportionality constant in \eqref{eq:pm-int} can be fixed by normalization. Note that $(\mathsf{v})^{\tau-1}$ means $\mathsf{v}$ applied $\tau-1$ times, while $k^{t}_{u_t,v_t}$ has a superscript. 
			
			\noindent\textbf{Numerical methods.} We discuss a few variants of \eqref{eq:pm-int} and their application in numerics. 
			
			When no measurement is performed at $t \notin \mathbf{t}$, we simply remove the integral over $u_t, v_t$, and $k^{t}_{u,v}$ in \eqref{eq:hatp-recurs} for $t \notin \mathbf{t}$. For example, if $\mathbf{t} = \{T\}$,  we have a considerable simplification: 
			\begin{equation} \label{eq:pt}
				Q^{\mathbf{t}}_{n_T} = \int d w e^{- \Gamma^2 w^2 / 2 - i w n_T}  (\mathsf{v})^T [e^{i w Z / \eta_T}] \,.
			\end{equation}
			Above we have denoted $w = u_T - v_T$ and integrated out $u_T + v_T$ on which $		Q^{\mathbf{t}}_{n_T} $ has no dependence. Taking the trace of this formula allows exact numerical calculation of $ p^{\mathbf{t}} $ in Fig.~\ref{fig:Delta}. Keeping all the components of $	Q^{\mathbf{t}}_{n_T} $, we obtain the pointer states ensemble from a single-time measurement, presented in Fig.~\ref{fig:pointer}. We have checked numerically that $Q^{\mathbf{t}}_{n_T}  $ has a well-defined $\Gamma \to 0, T \to \infty$ limit, from which the results in Fig.~\ref{fig:pointer} are indistinguishable.
			
			To compute a marginal distribution (and the associated pointer state), we integrate out $m_t$ for all $t \in \mathbf{t}$. This enforces $u_t = v_t =: z_t/2$. Therefore the $u_t, v_t$ integral in \eqref{eq:pm-int} is replaced by $\int_{z_t} e^{-  \Gamma z_t^2 / 2  } $ and $k^t_{u_t, v_t} $ in \eqref{eq:hatp-recurs} is replaced by $k^t_{z_t/2, z_t/2}$. For example, the marginal distribution (denoted by the superscript ``$\text{mar.}$'' below) to $\mathbf{t} = \{T\}$ is given by the following:
			\begin{align} \label{eq:marginal}
				p^{\mathbf{t}, \text{mar.}}_{n_T}	= & \mathrm{tr} (Q^{\{T\}, \text{mar.}}_{n_T}) / 2  \,, \text{where }  \\
				Q^{\mathbf{t}, \text{mar.}}_{n_T} =& \int_{w, \vec{z}} e^{- \Gamma^2 w^2 / 2 - i w n_T -  \sum_{t=\tau}^{T-1} \Gamma^2 z_t^2 / 2 }   \label{eq:Qmarginal} \\ 
				& (\mathsf{v})^\tau k^{\tau}_{\frac{z_\tau}2,\frac{z_\tau}2} \mathsf{v} \dots  \mathsf{v} k^{T-1}_{\frac{z_{T-1}}2,\frac{z_{T-1}}2} \mathsf{v}[e^{i w Z / \eta_T}]  \,,\nonumber
			\end{align}
			where $\vec{z} = (z_\tau, \dots, z_{T-1})$. This formula can be numerically estimated as follows. For a range of $w$, we directly sample the integral over $\vec{z}$ with the Gaussian weight $e^{-\sum \Gamma^2 z_t^2 / 2}$, and calculate the integrand (which involves manipulating a $2\times2$ matrix for $O(T)$ times). Then we perform the $w$ integral using fast Fourier transform. We observe excellent convergence of the estimate with $\sim 1000$ samples in all computations involved in Fig.~\ref{fig:Delta} (the whole figure takes few minutes on a laptop).
			
			\noindent\textbf{Linearized flow}. We next consider the behavior of $Q_t$ under the ``flow'' given by the backward recursion~\eqref{eq:hatp-recurs} at $t \gg 1$. Since the Gaussian weight in \eqref{eq:pm-int} restricts $u_t, v_t$ to be $ \sim O(1/\Gamma)$, $Q_t$ will be close to $I$. In fact, we may check inductively the following Ansatz for the leading behavior:
			\begin{equation}\label{eq:Qt-linear}
				Q_t - I = a_t  \eta_{t}^{-2} I + b_t \eta_t^{-1} O^{(1)} + O(\eta_t^{-2}) q + O(\eta_t^{-3})
			\end{equation}
			where $a_t, b_t \sim O(1)$ and $q$ is a combination of $O^{(1)}, O^{(2)}, O^{(3)}$ (but no identity). Indeed, the action of $k^t_{u_t, v_t}$ is 
			\begin{align}
				&	k^t_{u_t, v_t}[Q_t] - I  \label{eq:ktaction-linear}\\ 
				=& \left( a_t + i w_t b_t \cos\frac\theta2 - \frac{w_t^2}2 \right) \eta_t^{-2} I
				+ b'_t  \eta_t^{-1} O^{(1)}  + \dots  \nonumber 
			\end{align}
			where 
			\begin{equation} \label{eq:bprime}
				b'_t = b_t  + i w_t \frac{\cos\frac\theta2}{\cos\theta} \,,\, w_t = u_t - v_t  \,,
			\end{equation}
			and we omitted terms of higher order in $\eta_t^{-1}$, terms $\sim \eta_t^{-1}  O^{(2)} $  or $\sim \eta_t^{-1}  O_\iota$.  The last two terms can be ignored since they have $+\infty$ scaling dimension and can only survive $\mathsf{v}$ through OPE, generating higher order $O(\eta_t^{-2})$ terms. Then applying $\mathsf{v}$ we may obtain the linearized flow equations, exact in the $t \gg 1$ limit:
			\begin{subequations} \label{eq:recursion-linear}
				\begin{align} 
					& a_{t-1} =  \frac{\eta^2_{t-1}}{\eta^2_t} \left(  2 a_t + 2 i w_t b_t \cos\frac\theta2 + (b'_t)^2 - w_t^2 \right)  \\ 
					&b_{t-1} =  \frac{\eta_{t-1}}{\eta_t} 2 \cos\theta \, b'_t \,.
				\end{align}
			\end{subequations}
			They will be our main analytical tool below.
			
			\noindent\textbf{Argument for decoherent histories.} 
			We now argue that $\Delta_{\vec{n}}^{\textbf{t}} \to 0$ provided $ \tau \gg 1$. Qualitatively, the argument consists in noticing that in the linearized recursion relations~\eqref{eq:recursion-linear}, $u_t$ and $v_t$ appear only through the combination $w_t = u_t - v_t$. However, we have seen that, when computing a marginal distribution, we need to integrate over $m_t$ for $t\notin \mathbf{t}$, which enforces $w_t = u_t - v_t = 0$. Therefore the existence of the measurements at $t \notin \mathbf{t}$ has vanishing effect on the marginal distribution to $\mathbf{t}$. But this effect is precisely what is quantified by the probe $ \Delta_{\vec{n}}^{\textbf{t}} $, so we conclude that $ \Delta_{\vec{n}}^{\textbf{t}} $ will be vanishing.
			
			It is helpful to illustrate the argument in the case of $\mathbf{t} = \{T\}$ discussed above. We observe that the super-operators $k^t_{z_t/2, z_t/2}$ in \eqref{eq:marginal} can be removed since $u_t = v_t = z_t / 2 \implies w_t = 0$. But removing the $k^t_{z_t/2, z_t/2} $'s reduces \eqref{eq:marginal} to \eqref{eq:pt}; that is, as $\tau \gg 1$, the marginal distribution equals the distribution where measurements only happen at $t\in \mathbf{t}$: $Q^{\mathbf{t}, \text{mar.}} = Q^{\mathbf{t}}$ in the $\tau \gg 1$ limit.  
			
			To obtain a quantitative estimate of $ \Delta_{\vec{n}}^{\textbf{t}}$, we look for the leading $z_t$-depending term omitted in \eqref{eq:ktaction-linear} with $u_t = v_t = z_t / 2$: 
			\begin{align}
				&	k^t_{z_t/2, z_t/2}[Q_t]  \supset  e^{i z_t Z / 2\eta_t} b_t \eta_t^{-1} O^{(1)}  e^{ - i z_t Z / 2\eta_t} \nonumber \\ 
				\supset & \sin\frac{\theta}2 e^{i z_t Z / 2\eta_t} b_t \eta_t^{-1} Y  e^{ - i z_t Z / 2\eta_t}  \nonumber \\ 
				=& \sin\frac{\theta}2 b_t \eta_t^{-1} (Y + i z_t \eta_t^{-1} [Z, Y] -  z_t^2 \eta_t^{-2} [Z,[Z,Y]] / 8 + \dots)   \nonumber 
			\end{align} 
			Now, the first order term $ \sim [Z, Y] = -i X$ will suppressed by $\mathsf{v}$. Yet, the next term gives $[Z,[Z,Y]] = Y$ which has overlap with $O^{(1)}$ and should be kept. Hence, 
			\begin{equation}
				k^t_{z_t/2, z_t/2}[Q_t] - Q_t \sim 2  b_t  \eta_t^{-1} O^{(1)} \times   z_t^2 \eta_t^{-2} \,.
			\end{equation}
			Compared to \eqref{eq:ktaction-linear} and \eqref{eq:bprime} above, we see that $b'_t$ has the (subleading) correction $\delta b'_t  \sim   b_t  z_t^2 \eta_t^{-2} $. Propagating that through \eqref{eq:recursion-linear}, we get
			\begin{align} 
				& \delta b_{t-1} \sim b_t z_t^2 \eta_t^{-2} \,,\,  \delta a_{t-1} \sim b_t^2 z_t^2 \eta_t^{-2}
			\end{align}
			Since $b_t \sim O(1)$, and $z_t  \sim O(1/\Gamma)$, so we conclude that
			\begin{equation} \label{eq:Delta-general}
				\Delta_{\vec{n}}^{\mathbf{t}} = O((\Gamma \eta_\tau)^{-2})
			\end{equation}
			in general, at least for fixed $L = T - \tau$ as $\tau \to \infty$. Moreover, since $\eta_t$ decays exponentially in $t$, we expect late-time disturbance to be exponentially smaller. So we expect that  \eqref{eq:Delta-general} should hold uniformly in $L$.
			
			In the case of $\mathbf{t} = \{T\}$ that we studied numerically, $w_t = 0$ for all $ t < T$. Then it is not hard to see from \eqref{eq:recursion-linear} that 
			\begin{align}  \label{eq:bt-encoding-T}
				b_t  =  (2 \cos\theta)^{T-t}   \frac{\eta_t}{ \eta_T}  \, i w_T \frac{ \cos\frac\theta2 }{\cos\theta}
			\end{align} 
			When $x_1 < 1/2$, $b_t  \sim 1$ for all $t$, so \eqref{eq:Delta-general} is tight. When $x_1 >  1/2$, $b_t^2 \sim 2^{-(T-t)(2x_1-1)} \ll 1$ and $\delta a_{t-1} \sim 2^{T(1-2x_1)} 2^{ -2 t (1 - x_1)}$ [recall $\eta_t^2 \sim 2^t$ for $x_1 > 1/2$~\eqref{eq:etat_app}]. So the disturbance comes from $t \sim T$ when $x > 1$ and $t \sim \tau$ when $x < 1$ (disturbance propagating through $b_{t-1} O^{(1)}$ will have to come back to identity later in the flow, and be suppressed when $x_1 > 1/2$). We obtain the following scaling behaviors with respect to $L = T - \tau$ and $\tau$:
			\begin{equation}	\label{eq:decayrate}
				\Vert \Delta^{\mathbf{t}} \Vert_1  \sim \begin{cases}
					2^{-2\tau (1-x_1)}   &    x_1 < 1/2 \\
					2^{-\tau} 2^{L(1-2x_1)} & 1/2 < x_1 < 1 \\ 
					2^{-\tau - L}  & x_1 > 1 
				\end{cases}  \,.
			\end{equation}
			They are verified numerically in Fig.~\ref{fig:Delta}.

			\noindent\textbf{Statistics at $x_1 < 1/2$}. We consider the multi-time statistics and pointer states ensemble of the decoherent histories $\vec{m}$, assuming $\tau \gg 1$. To lighten the notation we will write 
			\begin{equation}
				s := \tau - 1 \gg 1\,.
			\end{equation}
			We first solve the linearized flow equation~\eqref{eq:recursion-linear} for $b_{\tau-1}$:
			\begin{align} \label{eq:bs-revealing}
				b_s  &= \frac{\cos\frac\theta2}{\cos\theta}  \sum_{t > s}   i w_t \frac{\eta_{s}}{\eta_t} (2 \cos\theta)^{t - s }  \nonumber \\ 
				&=\frac{\cos\frac\theta2}{\cos\theta}  \sum_{t > s} i w_t \,, 
			\end{align}
			where the last identity relies on the asymptotic law \eqref{eq:etat_app} and is exact in the $\tau \gg 1$ limit. Therefore, plugging back into \eqref{eq:pm-int}, we have 
			\begin{align} 
				Q_{\vec{m}} &= \int_{\vec{w}}   e^{ - \sum_t  (\Gamma^2 w_t^2 / 2  + i w_t {m_t} )}   (\mathsf{v})^{s}[Q])  \label{eq:pm-revealing1} \\ 
				Q & = I (1 + a_s \eta_s^{-2})+ O^{(1)} b_s \eta^{-1}_{s} \,.
			\end{align}
			Here we have used the independence on $z_t = u_t + v_t$ to reduce the $\vec{u}, \vec{v}$ integral to that of $\vec{w} = \vec{u} - \vec{v}$. Now recall that, in many-body terms, we have
			\begin{align}
				& (\mathsf{v})^{s}[Q]  =  \mathcal{V}_{0,s} [\prod_j Q_j] 
				=   \mathcal{V}_{0,s} [e^{\sum_j  (O_{x,j}  b_s \eta^{-1}_s  + O(\eta_s^{-2}) }]    \nonumber 
			\end{align}
			We see that the term $O(\eta_s^{-2})$ can be neglected, even if it involves the identity, since $\sum_j  \eta_s^{-2} = 2^{s} 2^{2s(x-1)} \ll1 $ for $x < 1/2$. Going to back to \eqref{eq:pm-revealing1}
			\begin{equation} \label{eq:Qm-apparatus}
				Q_{\vec{m}} = \int_{\vec{w}}   e^{ - \sum_t  (\Gamma^2 w_t^2 / 2  + i w_t {m_t} )}   (\mathsf{v})^{s}) [e^{O^{(1)} b_s / \eta_s}]
			\end{equation}
			where $b_s$ is given by \eqref{eq:bs-revealing}. Taking the rescaled trace of this formula,
			$$
			p_{\vec{m}} = \int_{\vec{w}}   e^{ - \sum_t  (\Gamma^2 w_t^2 / 2  + i w_t {m_t} )}   \mathrm{tr} ((\mathsf{v})^{s}[e^{O^{(1)} b_s / \eta_s}] )/2
			$$
			and recalling $b_s = \frac{\cos\frac\theta2}{\cos\theta}  \sum_{t > s} i w_t$, we see that $p_{\vec{m}}$ has the following law:
			\begin{equation}
				m_t \stackrel{\text{in law}}=   \mathcal{M} + \Gamma  f_t \,,   
			\end{equation}
			where $\mathcal{M}$ is a random variable with characteristic function 
			\begin{equation} \label{eq:bigM}
				\langle	e^{i w \mathcal{M} }  \rangle = \mathrm{tr}((\mathsf{v})^{s}[e^{i  \frac{\cos\frac\theta2}{\cos\theta} O^{(1)} w / \eta_s}]) / 2
			\end{equation}
			and $f_t$ are independent standard Gaussian variables. Since the marginal distribution of $m_t$ does not depend on $s$ as $s\gg1$, the right hand side of \eqref{eq:bigM} is independent of $s$ in the same limit. Comparing \eqref{eq:Qm-apparatus} to the marginal version \eqref{eq:marginal}-\eqref{eq:Qmarginal} we see that
			\begin{equation}
				(\mathsf{v})^{s}[e^{i  \frac{\cos\frac\theta2}{\cos\theta} O^{(1)} w / \eta_s}] = 	(\mathsf{v})^{s}[e^{i  Z w / \eta_s}] \,,\, s \to \infty \,.
			\end{equation}
			This means that $\mathcal{M}$ has the same distribution as $\mathcal{Q}_s / \eta_s = \sum_{j=1}^{2^s} Z_j / \eta_s$ in the $s \to \infty$ limit. In conclusion, a history instance $(m_t), t \ge \tau \gg 1$ is a Gaussian white noise of amplitude $\Gamma$ plus a time-independent random value whose distribution is the same as the rescaled coarse-grained observable in the long time limit.
			
			Concerning the pointer states ensemble, notice that if we take the $\Gamma \to 0$ limit (after the $\tau \to\infty$ limit) in \eqref{eq:Qm-apparatus}, we have, noting $q(w) := (\mathsf{v})^{T}(e^{i Z w / \eta_T})$, 
			\begin{align}
				\lim_{\Gamma \to 0}	Q_{\vec{m}}  &=  \int_{\vec{w}}   e^{ - \sum_t i w_t m_t }  q(\sum_t w_t) \nonumber  \\ 
				= &  \int_{\vec{w}}   e^{ -i \sum_t w_t m_T -  \sum_{t < T}  w_t (m_t - m_T) }  q(\sum_t w_t) \nonumber \\
				= & \prod_{t < T} \delta(m_t - m_T) \int_w  e^{ -i w m_T} q(w) \nonumber \\
				=& \lim_{\Gamma \to 0} Q^{\{T\}}_{m_T}  \prod_{t < T} \delta(m_t - m_T) 
			\end{align}
			where we compared to \eqref{eq:marginal}-\eqref{eq:Qmarginal} in the last line. Therefore the full history $\vec{m}$ infers no more information than a single snapshot $m_T$. A similar argument shows that for $\Gamma > 0$, the full history is equivalent to a snapshot with a reduced imprecision $\Gamma \to \Gamma / \sqrt{T - \tau}$. 
			
			\noindent\textbf{Statistics at $x > 1/2$}. To obtain the history statistics in the encoding phase we need to solve the linearized recursion \eqref{eq:recursion-linear} for $a_s$. When $x > 1/2$, the equation for $a_s$ simplifies 
			\begin{equation}
				a_s = \frac12 \sum_{t > s}  (  2 i w_t b_t \cos\frac\theta2  + (b'_t)^2- w_t^2 ) 
			\end{equation} 
			where 
			\begin{align}
				b_t &= i \frac{\cos\frac{\theta}2}{\cos\theta} \sum_{t' > t}  \lambda^{t'-t} w_{t'} , \lambda = \sqrt{2} \cos\theta \,, \\
				b'_t &= i  \frac{\cos\frac{\theta}2}{\cos\theta}  \sum_{t' \ge t} \lambda^{t'-t} w_{t'} = b_{t-1} \,.
			\end{align}
			After some arrangement we have 
			\begin{align}
				a_s =& - \frac12 \sum_{tt'} C_{tt'} w_t w_{t'}  \\ 
				C_{tt'} =& \delta_{tt'} + \frac{\cos^2\frac\theta2}{\cos^2\theta}  \sum_{r \le t, t'} \lambda^{t+t'-2r} \nonumber \\
				& +  \frac{ \cos^2\frac\theta2   }{\cos\theta} (1-\delta_{tt'}) \lambda^{|t-t'|}  \nonumber \\ 
				\sim &\lambda^{|t- t'|} \,,\, |t - t'| \gg 1 \,.  
			\end{align}
			To proceed we consider the marginal law of $m_t$ with $t \ge \tau'$ for some $\tau' - s \gg 1$, by setting $w_t = 0$ for $t < \tau'$. We can do this without loss of generality since $\tau'$ can take any large value just as $s$. Then $b_s \sim \lambda^{\tau' - s} \ll 1$ is small, and $Q_s = 1 + a_s \eta_s^{-2} = e^{ a_s \eta_s^{-2}} $ at leading order. Plugging back into \eqref{eq:pm-int}, after inverting the Fourier transform we have 
			\begin{align}
				& \int 	Q_{\vec{m}} e^{i \sum_{t \ge \tau'} m_t w_t}  =  e^{-\sum_t \Gamma^2 w_t^2 / 2}  (\mathsf{v})^{s} [ e^{ a_s \eta_s^{-2}} ]
			\end{align} 
			Now, recalling that $\eta_s^2 \sim c 2^{s}$~\eqref{eq:etat_app} with $c = 1 - \frac{\cos^2\frac\theta2}{\cos(2\theta)}$, we have 
			\begin{equation}
				\mathrm{tr}((\mathsf{v})^{s} [ e^{ a_s \eta_s^{-2}} ])   / 2
				=  e^{ \sum_{j = 1}^{2^s} a_s \eta_s^{-2}} = e^{a_s / c}
			\end{equation}
			We conclude that $(m_t)$ is a Gaussian process with zero mean and covariance matrix 
			\begin{align}
				\mathbb{E}[ m_t m_{t'} ] &= C_{tt'} / c+ \Gamma^2 \delta_{tt'}  \\ 
				& \sim \lambda^{|t-t'|} = (\sqrt{2} \cos\theta)^{-|t-t'|} = 2^{- |t-t'| (x_1 -1/2)} \nonumber
			\end{align}
			as $|t - t'| \gg 1$. 
			
			Since $ Q_{\vec{m}} $ is proportional to identity, we conclude that the conditional reduced density matrix of $S$ is $\rho_{\vec{m}} = I / 2$. So no information is inferred about the system from the full history. 
			
			\subsection{Leggett-Garg inequality}
\begin{figure}[h]
				\centering
				\includegraphics[scale=.9]{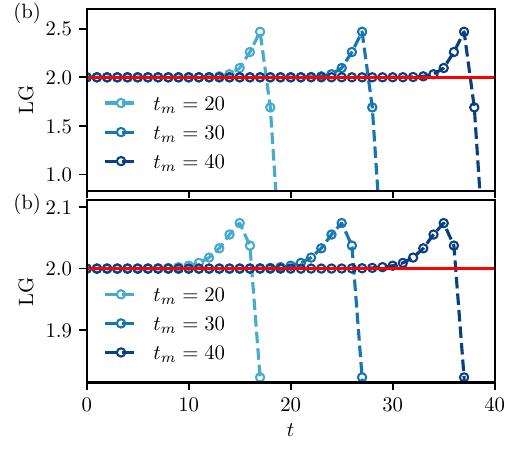} 
				\caption{The violation of the Leggett-Garg inequality in the apparatus phase (a, $\theta = 0.15 \pi$), and in the encoding phase (b, $\theta = 0.35\pi$), for $t_1, \dots, t_4 = t, \dots, t+3$, and for the operators defined by \eqref{eq:Qt-def}, \eqref{eq:abc-def}. The violation, indicated by the data points going above the red line, can happen at arbitrarily late time, controlled by the parameter $t_m$. } \label{fig:LG}
			\end{figure}

			In this section we show numerically the violation of the Leggett-Garg inequality in our model, with respect to the operators 
			\begin{equation}\label{eq:Qt-def}
				Q_t =  (a X + b Y + c Z)^{\otimes 2^t} \,,\, a^2 + b^2 + c^2 = 1 \,,
			\end{equation}
			so that $Q_t^2 = \mathbf{1}$; that is, the eigenvalues of $Q_t$ are $\pm 1$. In our inflationary model, a two-time Keldysh correlation function should be defined as 
			\begin{equation} \label{eq:Cst}
				C_{st} = \mathrm{Re} \left<   V_{t,0}^\dagger Q_s V^\dagger_{t, s} Q_t V_{t, 0}\right>  \,, s < t \,.
			\end{equation}
			Then the Leggett-Garg inequality is the statement that for any  $t_1 < t_2 < t_3 < t_4$,
			\begin{equation}
				\mathrm{LG} :=	C_{t_1 t_2} + 	C_{t_2 t_3} + C_{t_3 t_4} - C_{t_1 t_4} \le 2 \,.
			\end{equation}
			The inequality holds if the correlation functions are between $4$ classical variables taking values in $\pm 1$, but can be violated when the correlation functions are Keldysh correlators between operators with eigenvalues $\pm1$. The correlation function~\eqref{eq:Cst} can be evaluated exactly and efficiently in our model using a similar operator method as above, since $Q_t$ is a product operator. In fact, in terms of $\mathsf{v}$ defined in \eqref{eq:vk}, we have 
			\begin{equation}
				C_{st} = \mathrm{Re} \,  \mathrm{tr}(   (\mathsf{v}^{s}) [Q  \mathsf{v}^{t-s}[Q]] )  \,,\,  Q = a X + b Y + c Z \,.
			\end{equation}	
			In Fig.~\eqref{fig:LG} we show that Leggett-Garg inequality is violated by the following choice:
			\begin{align} \label{eq:abc-def}
				&(a, b, c)=  (\cos(\epsilon) , \sin(\epsilon) \cos(\theta/2), \sin(\epsilon) \sin(\theta/2) ) \,,\, \nonumber \\
				&\epsilon = 1  / \max(\sqrt{2}, 2 \cos\theta  )^{t_m}
			\end{align}
			By tuning $t_m$ we can let the violation take place at arbitrarily late time $t \sim t_m$. 
			
			We speculate that the violation is related to the $\mathbb{Z}_2$ symmetry of the model, 
			\begin{equation}
				V_{t+1,t} X^{\otimes 2^{t}} = X^{\otimes 2^{t+1}}  V_{t+1,t}  \,.
			\end{equation}
			Indeed, the operators defined by \eqref{eq:Qt-def} and \eqref{eq:abc-def} are close to the symmetry operator. The precise relation between non-classicality and quantum symmetry is left to future study.

\end{document}